\input{aipcheck}

\documentclass[
    ,final            
  ]
  {aipproc}

\layoutstyle{6x9}

\begin{document}

\title{The near-threshold peak of the $\phi$-photoproduction cross section \\ and its interpretation}

\classification{13.60.Le, 25.20.Lj, 14.20.Gk}
\keywords      {Photoproduction, $\phi$ meson, nucleon resonance, Pomeron}

\author{Alvin Kiswandhi}{
  address={Department of Physics and Center for Theoretical Sciences,\\ National Taiwan University,
Taipei 10617, Taiwan}
}

\author{Shin Nan Yang}{
}


\begin{abstract}

We study, within phenomenological Lagrangian approach, the possibility that the near-threshold peak 
found in the $\phi$-photoproduction cross section is caused by a resonance. We show that, by employing a $J^P=3/2^-$ resonance with a mass of
$2.08\pm 0.05$ GeV and a width of $ 0.570\pm 0.169$ GeV, the LEPS data which include new data on nine spin-density matrix elements 
can indeed be described reasonably well. We also find that the ratio of helicity amplitudes $A_{1/2}/A_{3/2}$ calculated from the resulting coupling
constants differs in sign from that of $D_{13}(2080)$. The resonance is further found to be able to improve the theoretical description of $\omega$ photoproduction 
against a new set of data if a large value for OZI evading parameter is assumed.

\end{abstract}

\maketitle




The appearance of a local maximum at around
$E_{\gamma}\sim 2.0$ GeV in the differential cross section (DCS)
of $\phi$ photoproduction on protons at forward angles has been
known for some time \cite{leps05}. However, it has been found that
it is not possible to describe the peak by $t$-channel
exchanges only \cite{titov97}-\cite{titov07}. 



Here, we follow from Ref. \cite{kiswandhi10}, where we study whether the nonmonotonic behavior found in
Ref. \cite{leps05} can be described by a resonance. Namely, we will add a
resonance to a model consisting of Pomeron and $(\pi,\eta)$ exchanges \cite{bauer78,donnachie87}, 
and see if we can better describe the experimental data
by a suitable set of properties of the resonance which include spin, parity, mass, width and coupling constants.
The closeness of the resonance to the threshold suggests that its spin would probably just either be $1/2$ or $3/2$.
Similar analysis was also done in a coupled-channel model in Ref. \cite{ozaki09}. However, the analysis was redone
to correct a previous confusion in the phase of the Pomeron-exchange amplitude \cite{hosaka10}.

However, we improve upon our previous study \cite{kiswandhi10} by fitting our model to a new 
LEPS spin-density matrix elements (SDME) data \cite{leps10}, in addition to the previous DCS data \cite{leps05,durham}. 



\begin{figure}[t]

\includegraphics*[height=0.22\textheight,angle=0]{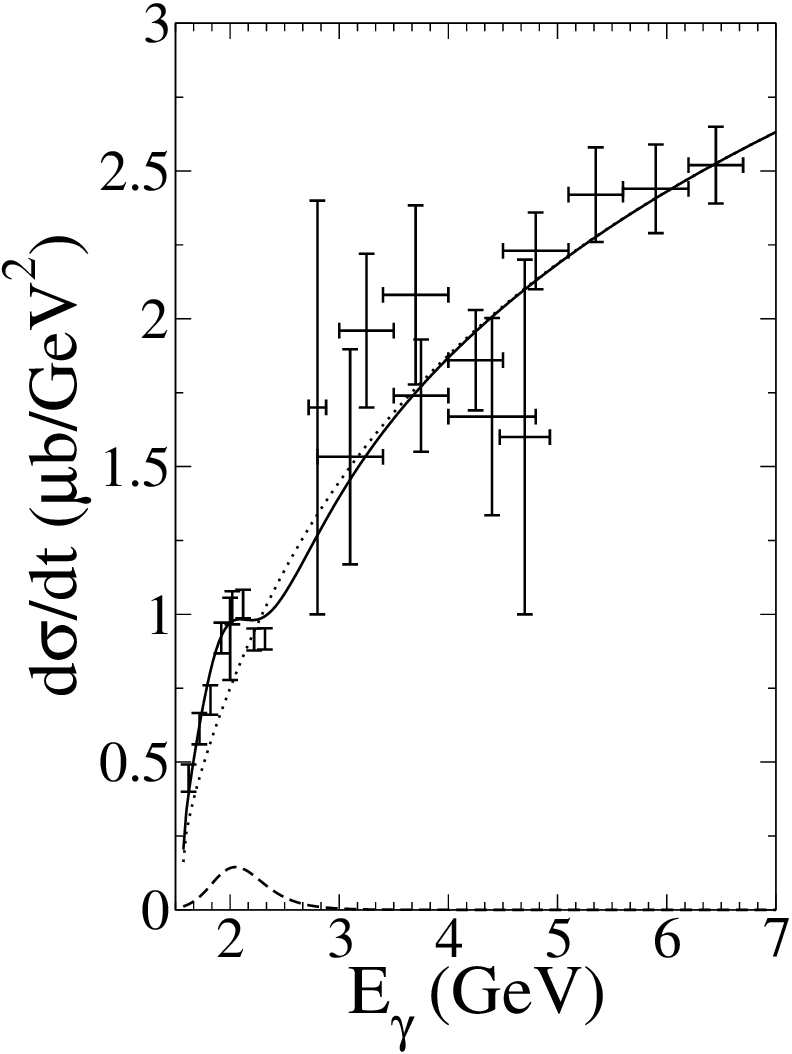}

\hspace{-0.3cm}

\includegraphics*[height=0.23\textheight,angle=0]{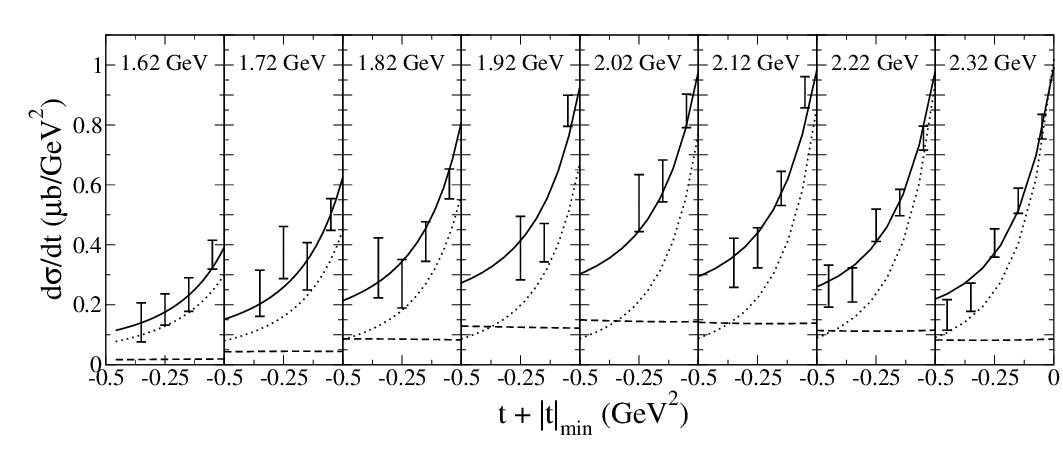}

\caption{Our results for the DCS of $\gamma p \to \phi p$ at
forward direction as a function of photon energy $E_\gamma$ (left) and as a function of $t$
at eight different photon LAB energies (right). Data are from Refs.~\cite{leps05,durham}. The
dotted, dashed, and solid lines denote contributions from nonresonant, resonance with $J^P = 3/2^-$, and their sum,
respectively.
\vspace{-0.75cm} }
\label{D13_1}
\end{figure}

The tree-level invariant amplitudes used here are given in Ref. \cite{kiswandhi10}. 
Only the mass, width, and the products of coupling constants of the resonance are free parameters and they are determined by the use of MINUIT,
by fitting to the LEPS as well as previous experimental data \cite{leps05, leps10, durham}.

We found that the nonmonotonic behavior of the DCS at forward direction as a function
of photon energy cannot be describe by the nonresonant contribution only. Moreover, an addition of a $J^P = 1/2^\pm$ resonance also cannot produce such behavior
near threshold, in contrast to the finding of Refs. \cite{ozaki09,hosaka10}. 

We found that both $J^P = 3/2^\pm$ resonances can describe the data reasonably well.
However, the extracted properties of the $J^P = 3/2^-$
resonance are more stable against the use of different data sets used in our previous study \cite{kiswandhi10}, compared to that of $J^P = 3/2^+$.
Therefore, we prefer the choice of $J^P = 3/2^-$ resonance with mass and width of $2.08 \pm 0.05$ and $0.570 \pm  0.169$, respectively.
The resulting coupling constants are 
$eg_{\gamma N N^*}^{(1)}g_{\phi N N^*}^{(1)} = -0.205 \pm  0.095$, 
$eg_{\gamma N N^*}^{(1)}g_{\phi N N^*}^{(2)} = -0.025 \pm  0.017$, 
$eg_{\gamma N N^*}^{(1)}g_{\phi N N^*}^{(3)} = -0.033 \pm  0.018$, 
$eg_{\gamma N N^*}^{(2)}g_{\phi N N^*}^{(1)} = -0.266 \pm  0.136$, 
$eg_{\gamma N N^*}^{(2)}g_{\phi N N^*}^{(2)} = -0.033 \pm  0.033$, 
and $eg_{\gamma N N^*}^{(2)}g_{\phi N N^*}^{(3)} = -0.043 \pm  0.032$.




Our best fits with $J^P = 3/2^-$ to the experimental
energy dependence of the DCS at forward angle and angular dependence of the DCS
\cite{leps05,durham} are shown in Fig.~\ref{D13_1}. 
One sees from 
Fig.~\ref{D13_1} that the resonance improves the agreement with the data

\begin{figure}[htbp]

\includegraphics[height=0.29\textheight,angle=0]{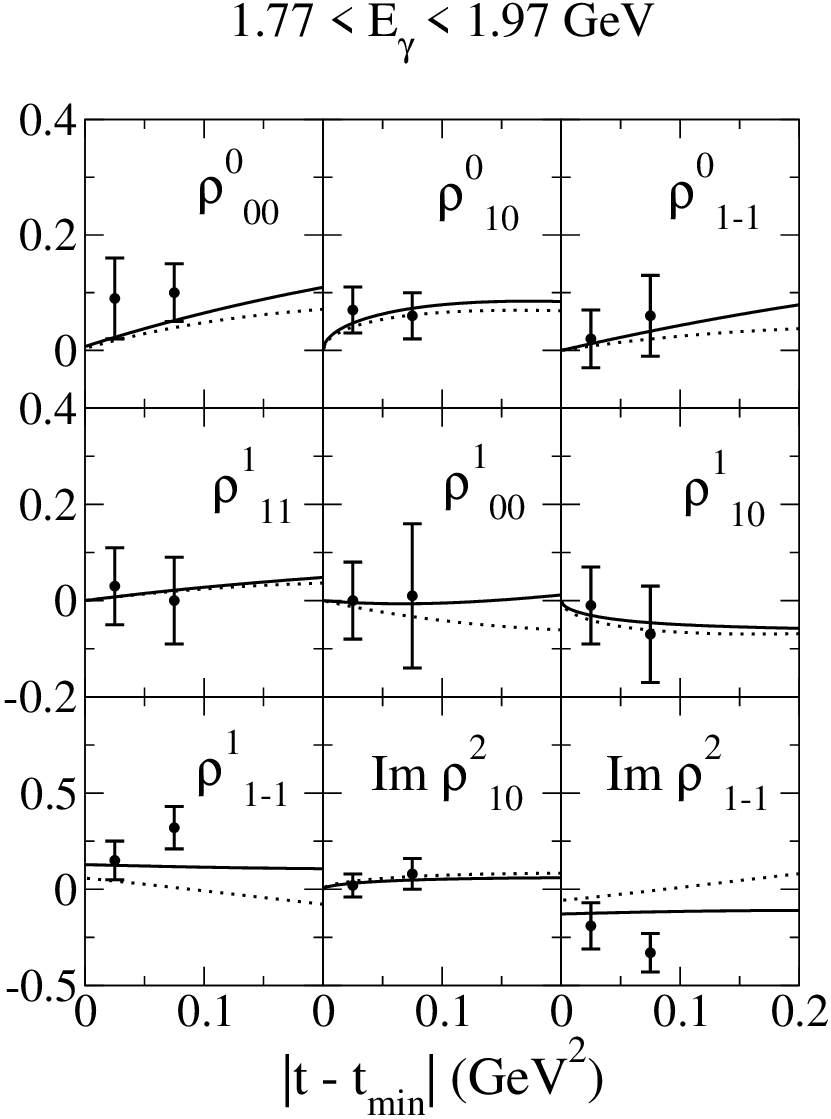}

\hspace{0.1cm}

\includegraphics[height=0.29\textheight,angle=0]{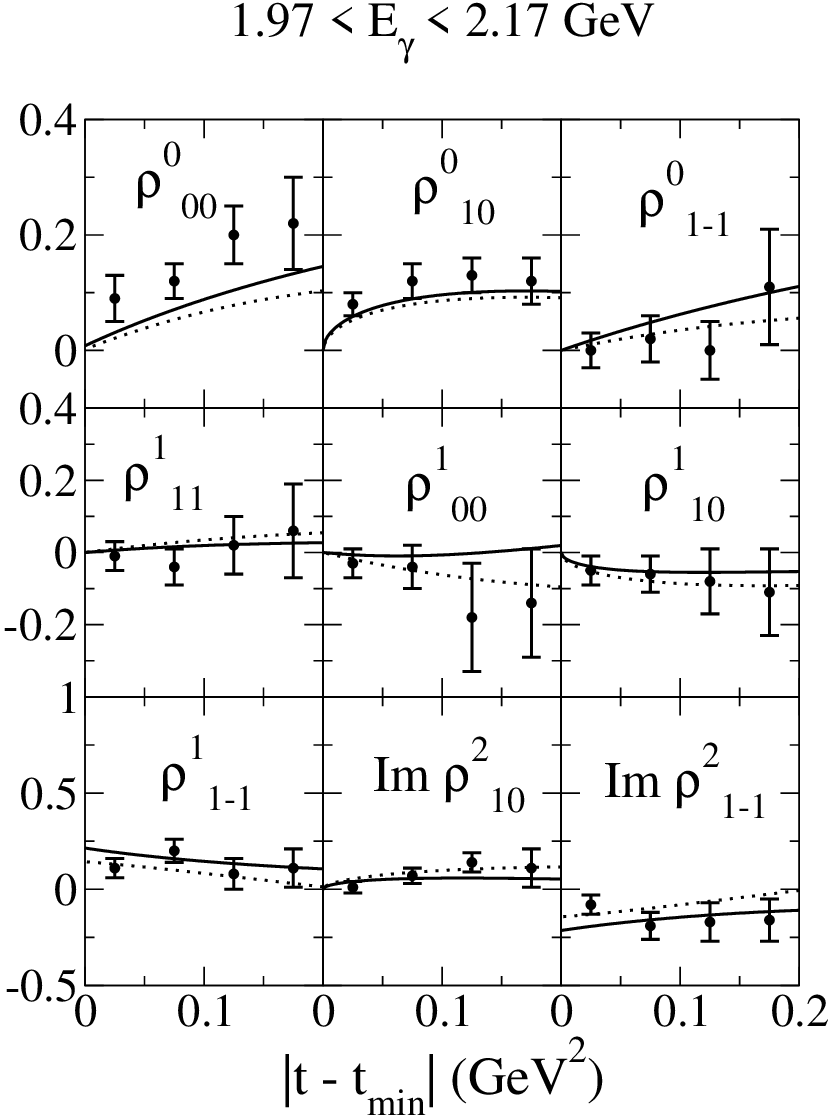}

\hspace{0.1cm}

\includegraphics[height=0.29\textheight,angle=0]{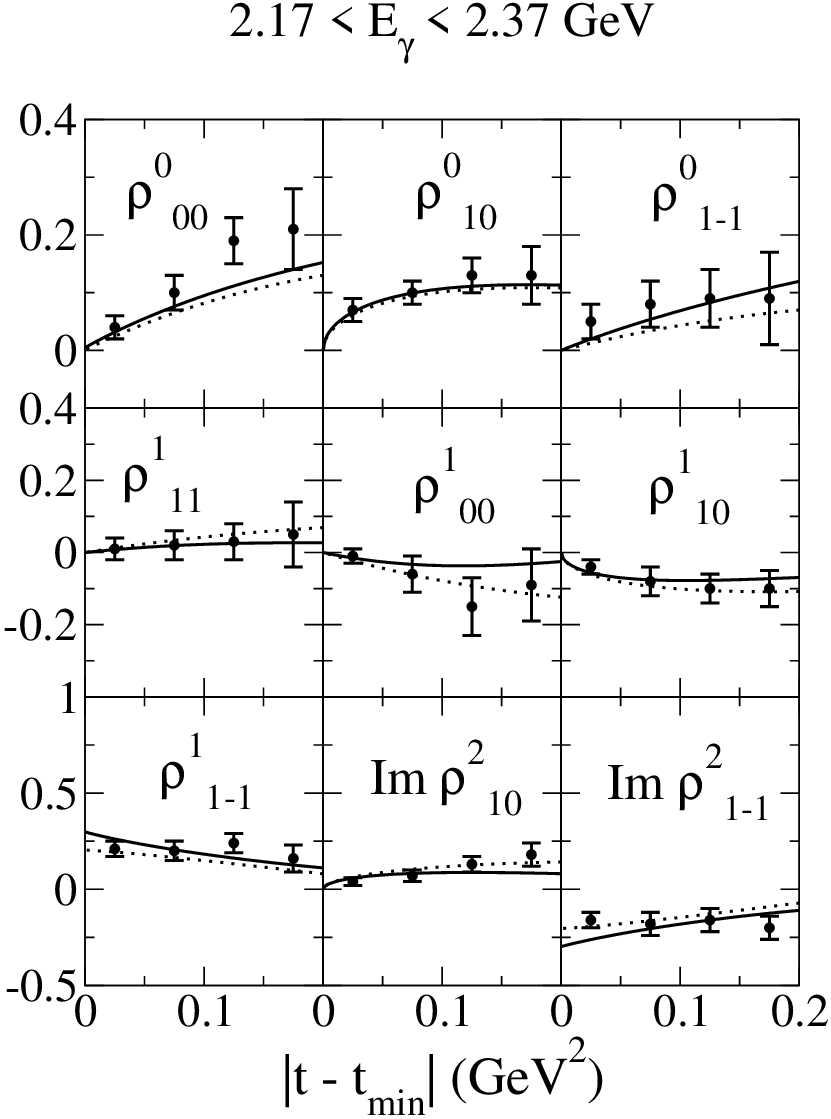}

\caption{{Our results for SDME in Gottfried-Jackson system obtained with  $J^P = 3/2^-$ resonance at three photon LAB energies, from left to right, $1.77 - 1.97$ GeV,
$1.97 - 2.17$ GeV, and $2.17 - 2.37$ GeV. Data is taken from Ref.~\cite{leps10}. The notation is the same as in Fig.~\ref{D13_1}. \vspace{-0.cm}}}
 \label{D13_3}
\end{figure}
Our results for the SDME in the Gottfried-Jackson system ~\cite{titov03,schillingnpb15397}, are shown in
Fig. \ref{D13_3}. Here, the inclusion of resonant contribution does help the agreement with the data in some cases, 
especially for $\rho^1_{1-1}$ and $\textrm{Im}\rho^2_{1-1}$ at $1.77-1.97$ GeV. However, for $\rho^1_{00}$ at all energies,
it does actually reduce the agreement, already good without the inclusion of the resonance, although still reasonably within error bars.
For other $\rho$'s, the effects are somewhat minimal, which is expected since the $t$-channel exchanges results are already good.


We caution against an attempt to identify the $3/2^-$ as the $D_{13}(2080)$
as listed in PDG \cite{PDG02}. With the coupling constants given above, we obtain a value of
{$A_{1/2}/A_{3/2}=1.05$} which differ from $-1.18$ for $D_{13}(2080)$ \cite{PDG02} in relative sign.

\begin{figure}[b]

\includegraphics[height=0.25\textheight,angle=0]{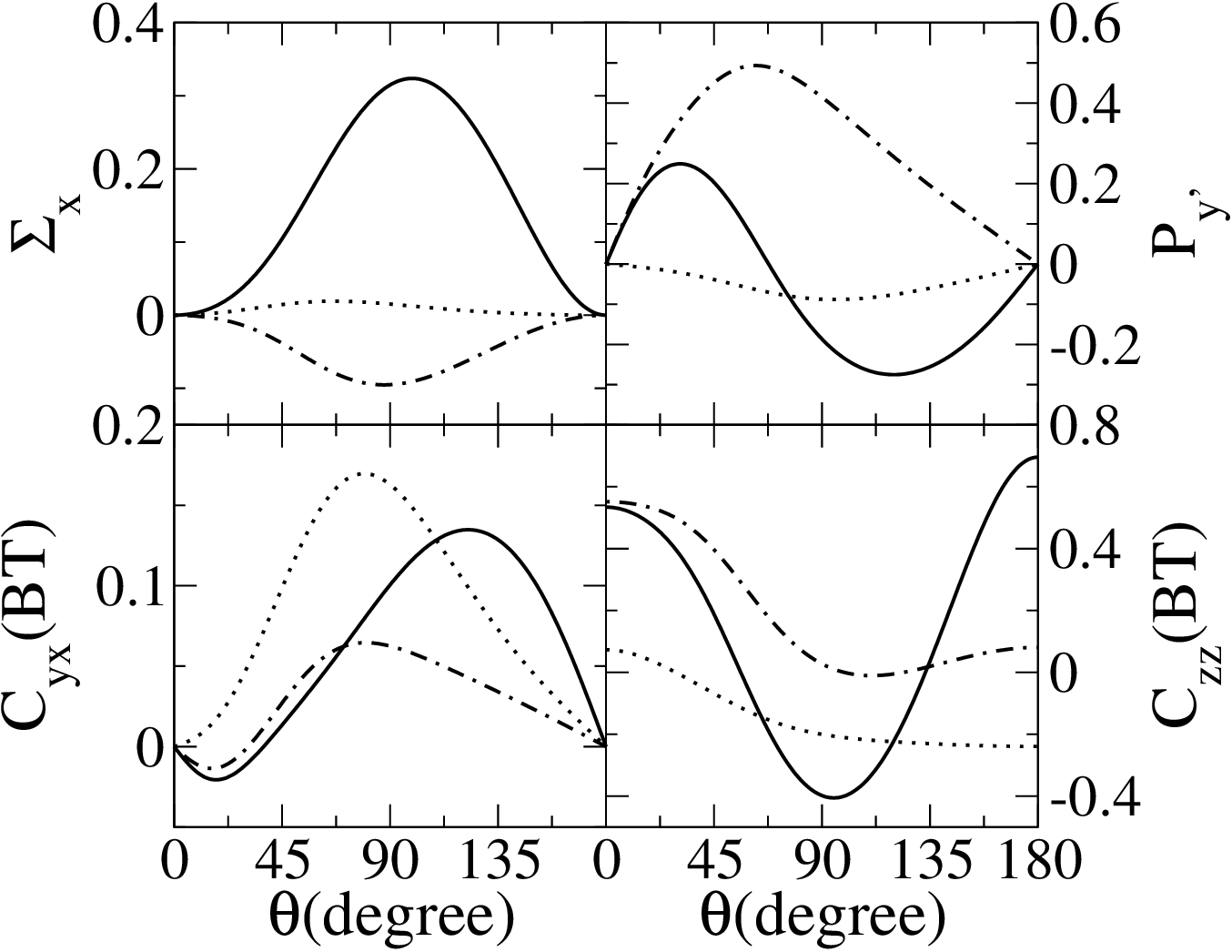}

\includegraphics[height=0.25\textheight,angle=0]{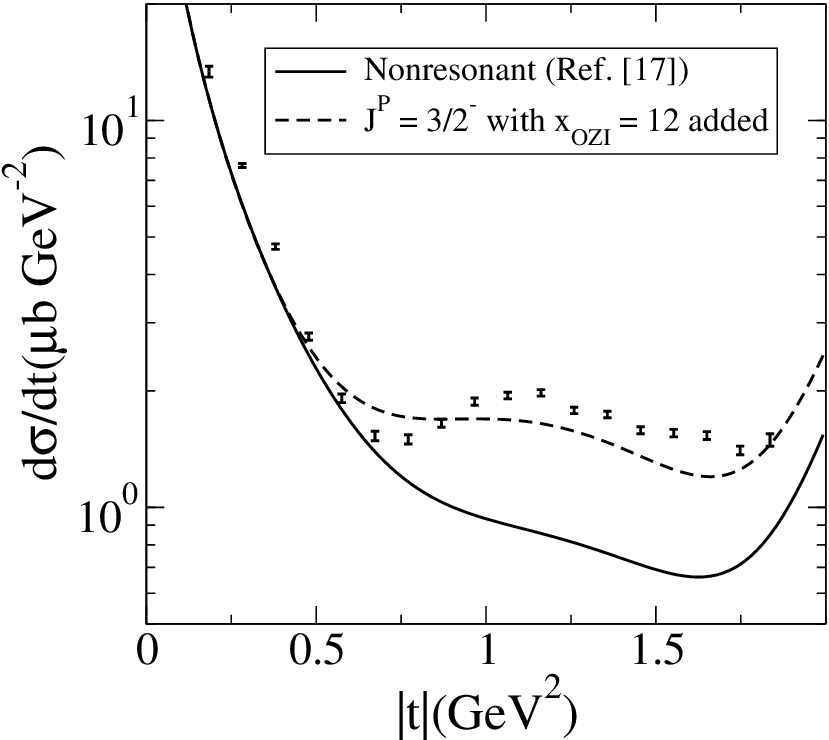}

\caption{{Left: Single and double polarization observables
$\Sigma_x$, $P_{y'}$, $C_{yx}^{BT}$, and $C_{zz}^{BT}$ taken at photon
laboratory energy $E_\gamma = 2$ GeV. The solid and dash-dotted
lines correspond to our results with the choices of $J^P = 3/2^-$ and
$J^P = 3/2^+$, respectively, while the dotted lines denote the
nonresonant contribution. Right: DCS of $\omega$ photoproduction as a
function of $|t|$ at $W = 2.085$ GeV. Solid and dashed lines
represent the model predictions of Ref. \cite{oh02} without and with
the addition of our preferred $J^P = 3/2^-$ resonance with $x_{\mathrm{OZI}} = 12$. Data are
from Ref. \cite{M_Williams}.}} \label{Polarization}

\end{figure}

We also find that the effects of the resonance are generally quite substantial in many polarization observables {\cite{titov_polarization}}. Results for some  
single and double polarization observables $\Sigma_x$, $P_{y'}$, $C_{yx}^{BT}$, and $C_{zz}^{BT}$ are shown in left panel of Fig.~\ref{Polarization}. 
The results using a $3/2^+$ resonance are also shown by dash-dotted curve to show that the measurements of these observables would help to determine 
the parity of the resonance.



We also expect that a resonance in $\phi N$ channel would also appear 
in the $\omega N$ channel because of the $\phi-\omega$ mixing
The conventional "minimal" parametrization relating $\phi NN^*$ and
 $\omega NN^*$ is $g_{\phi N N^*} = -\tan \Delta \theta_V
x_{\mathrm{OZI}} g_{\omega N N^*}$, with $\Delta \theta_V \simeq 3.7^\circ$ corresponds to the
deviation from the ideal $\phi-\omega$ mixing angle.
The larger the value of the OZI-evading parameter $x_{\mathrm{OZI}}$, the larger is the strangeness
content of the  resonance.

Here, we add the resonance postulated here to the model of Ref. \cite{oh02} with $
x_{\mathrm{OZI}}=12$, whose prediction is given in the dashed line in right panel of 
Fig.~\ref{Polarization}. We see that the DCS at $W = 2.085$ GeV can be
reproduced with roughly the correct strength. However, it is shown in Ref. \cite{kiswandhi10} that this value of $x_{\mathrm{OZI}}$
produces better agreement to the data at $W = 2.105$ GeV. The large value of
$x_{\mathrm{OZI}}=12$ would imply that the resonance would contain 
a considerable amount of strangeness content.

In summary, we study the possibility that the near-threshold nonmonotonic behavior
of $\phi$-photoproduction cross section observed by the LEPS collaboration as a 
possible signature of a resonance. We confirm that this behavior cannot be explained by 
the nonresonant contribution alone, as well as by adding a resonance with $J = 1/2$. 
However, a resonance with $J^P=3/2^-$ would bring a reasonable agreement with most of the LEPS data, 
with a greater stability with respect to changes of the new data used in the fitting \cite{leps10}, compared to $J^P=3/2^+$.
The resonance mass and width obtained are
$2.08\pm 0.05$ and $0.570\pm 0.169$ GeV, respectively.  The ratio of the helicity amplitudes calculated from the coupling constants
differs from that of the known $D_{13}(2080)$ by a minus sign.
We also find that the resonance contributes significantly to the polarization observables, which can also be used
to determine the parity of the resonance if it indeed exists.
When a $J=3/2^-$ resonance is added to the model of Ref. \cite{oh02} with a large value of OZI-evading
parameter $x_{\mathrm{OZI}}=12$, the agreement of the model prediction with the most recent data is improved substantially,
which implies that the resonance would contain a considerable amount of strangeness content.



\begin{theacknowledgments}
We would like to thank Profs. W. C. Chang, C. W. Kao, A. Hosaka,
T.-S. H. Lee, Y.S. Oh, S. Ozaki, and A. I. Titov, for useful
discussions and/or correspondences. Travel grant from the National Center for Theoretical Sciences, Taipei, Taiwan, R.O.C. is also greatly appreciated. We also acknowledge the help from NTU HPC Center. This work is supported in part by the NSC of R.O.C. (Taiwan) 
under grant NSC99-2112-M002-011. 
\end{theacknowledgments}



\bibliographystyle{aipproc}   

\bibliography{sample}

\end{document}